\documentclass[aps,twocolumn,prd,nofootinbib,superscriptaddress,preprintnumbers,10pt]{revtex4-1} 

\usepackage{amsmath,amssymb,amsthm}
\usepackage{graphicx}
\usepackage{xcolor}
\usepackage{dcolumn}
\usepackage[breaklinks=true]{hyperref}
\usepackage{ascmac} 
\newcommand{\ds}{\displaystyle}
\newcommand{\tf}{\tfrac}
\newcommand{\vev}[1]{\left\langle #1 \right\rangle}
\newcommand{\vevs}[1]{\langle #1 \rangle}

\newcommand{\link}{\,{\rm link}\,}
\newcommand{\er}[1]{Eq.~\eqref{#1}}

\newcommand{\bb}{\mathbb}

\newcommand{\df}{\dfrac}
\newcommand{\fr}{\frac}

\newcommand{\der}{\partial}

\renewcommand{\(}{\left(}
\renewcommand{\)}{\right)}

\newcommand{\wed}{\wedge}
\newcommand{\bmx}{\left(\begin{matrix}}
\newcommand{\emx}{\end{matrix}\right)}

\begin{document}

\preprint{RIKEN-QHP-433, KEK-TH-2176}

\title{Emergent discrete 3-form symmetry and domain walls}

\author{Yoshimasa Hidaka}
\email[]{hidaka@riken.jp}
\affiliation{Nishina Center, RIKEN, Wako 351-0198, Japan}
\affiliation{RIKEN iTHEMS, RIKEN, Wako 351-0198, Japan}

\author{Muneto Nitta}
\email[]{nitta@phys-h.keio.ac.jp}
\affiliation{Department of Physics \& Research and Education Center for
Natural Sciences, Keio University, Hiyoshi 4-1-1, Yokohama, Kanagawa
223-8521, Japan}

\author{Ryo Yokokura}
\email[]{ryokokur@post.kek.jp}
\affiliation{KEK Theory Center, Tsukuba 305-0801, Japan}
\affiliation{Department of Physics \& Research and Education Center for
Natural Sciences, Keio University, Hiyoshi 4-1-1, Yokohama, Kanagawa
223-8521, Japan}

\date{\today}

\begin{abstract}
We show that axion models with the domain wall number $k$ in $(3+1)$
dimensions, {\it i.e.}, periodic scalar field theories admitting
$k$ axion domain walls, exhibit 
an emergent $\bb{Z}_k$ 3-form symmetry for $k >1$
in addition to a conventional ${\mathbb Z}_k$ 0-form symmetry.
The emergent 3-form symmetry is explicitly shown by establishing a
low-energy dual transformation between the scalar field theory and a
3-form gauge theory.
We further argue that the emergent 3-form symmetry is spontaneously
broken, and the breaking pattern is so-called the type-B spontaneous
symmetry breaking.
We discuss similar and different points between the phase
admitting the domain walls and topologically ordered phases.
\end{abstract}

\maketitle

\section{Introduction}
Classifying states of matter is one of the most important problems in
modern physics.
Spontaneous symmetry breaking separating ordered and disordered phases
is one key ingredient for that purpose, and the Ginzburg-Landau theory
based on local order operators offers a ubiquitous tool.
For the Ginzburg-Landau theory, the symmetries act on local
operators such as fields at points in spacetime.

Recently, a more general notion, higher-form symmetries, was
proposed~\cite{Banks:2010zn,Kapustin:2014gua,Gaiotto:2014kfa} (see
earlier references~\cite{Pantev:2005rh,Pantev:2005zs,Pantev:2005wj} and
related topics~\cite{Hellerman:2010fv}).
The higher $p$-form symmetries are symmetries under transformations of
$p$-dimensional non-local operators such as Wilson loops ($p=1$), world
surface of vortices ($p=2$), and so on.
In terms of the higher-form symmetry, more general phases can be
classified beyond the Ginzburg-Landau theory.

As applications, phases admitting topological solitons can be
 classified in terms of higher-form symmetries associated to the
 solitons.
Here, topological solitons are classical solutions in field theories
that have topological charges.
For example, fractional quantum Hall states in $(2+1)$ dimensions and
s-wave superconductors in $(3+1)$ dimensions are identified as broken
phases of discrete $p$-form and $(D-p-1)$-form
symmetries~\cite{Banks:2010zn,Kapustin:2014gua,Gaiotto:2014kfa} for
$(D,p) = (3,1)$ and $(4,1)$, respectively.
For the s-wave superconductors, the extended objects are a 1-form
worldline and a 2-form worldsheet, which represent the trajectory of an
electron and an Abrikosov--Nielsen--Olesen
vortex~\cite{Abrikosov:1956sx,Nielsen:1973cs}, respectively.
While both of them have been known as topologically ordered phases
because the extended objects lead to non-trivial braiding
statistics~\cite{Wen:1989iv,Wen:1990zza,Wen:1991rp,Hansson:2004wca},
they can now be understood in terms of the spontaneous symmetry
breaking.

Once the above two cases can be understood in terms of the symmetry,
they can be further classified as so-called type-B spontaneous
symmetry breaking~\cite{Watanabe:2011ec,Watanabe:2012hr,Hidaka:2012ym,
Watanabe:2014fva, Hayata:2014yga,Takahashi:2014vua,Watanabe:2019xul},
since the charged object that is the order operator is also the symmetry
generator.
In particular, the low-energy effective theories are written in the
topological quantum field theories with first-order temporal derivative
terms.
For s-wave superconductors, a $BF$ theory~\cite{Horowitz:1989ng,Blau:1989bq} is a
low-energy effective theory, which is obtained by a dual
transformation of an Abelian-Higgs model with magnetic vortices.
In the $BF$ theory, a 2-form gauge field couples to the magnetic
vortices~\cite{Sugamoto:1978ft,Lee:1993ty} (see also
Ref.~\cite{Nitta:2018qqe} as a recent reference).
We can see that a Wilson loop and a vortex world sheet can have a
nontrivial linking phase.
A generalization to non-Abelian (color) superconductors has been
recently discussed \cite{Hirono:2018fjr,Hirono:2019oup,Hidaka:2019jtv}. 

One of the natural questions may be whether there exists an ordered
phase characterized by domain walls instead of vortices in the s-wave
superconductor \footnote{In a spin system, a topologically ordered phase
due to the domain wall condensation has been
discussed~\cite{Hamma:2004ud}.}.
Domain walls appear in many contexts
in physics, e.g., magnetic domain walls in condensed matter, axionic
domain walls in cosmology, branes in string theory.
They can be topological solitons.
Domain walls in the $\Phi^4$ and sine-Gordon models are typical examples.

In this paper, we focus on axion models with the
domain wall number $k$ that are periodic scalar field theories
admitting $k$ axion domain walls in $(3+1)$ dimensions.
We show that there are non-local order parameters given by a domain wall
worldvolume and two points where the scalar field is locally put.
They develop non-zero vacuum expectation values (VEVs).
We find that the correlation function of the order parameters has a
fractional phase if the two points are separated by the domain wall, that
is, the domain wall worldvolume and the two-point operators are
linked.

The correlation function can be evaluated by using a dual 3-form
gauge theory.
Such 3-form gauge theories in $(3+1)$ dimensions have been
considered in many contexts, e.g., quantum
chromodynamics~\cite{Aurilia:1977jz,Aurilia:1978dw,Luscher:1978rn,Aurilia:1980xj,Aurilia:1980jz,Hata:1980hn},
the strong CP problem~\cite{Dvali:2005zk,Kaloper:2017fsa}, the
cosmological constant
problem~\cite{Aurilia:1980xj,Hawking:1984hk,Brown:1987dd,Brown:1988kg,Duff:1989ah,Duncan:1989ug,Duncan:1990fr},
inflationary
models~\cite{Kaloper:2008fb,Kaloper:2008qs,Kaloper:2011jz,Marchesano:2014mla,Dudas:2014pva,Kaloper:2016fbr,Kaloper:2017fsa,DAmico:2017cda,DAmico:2018mnx,Yamada:2018nsk},
string effective theories~\cite{
Ibanez:2014swa,Bielleman:2015ina,Ibanez:2015fcv,Valenzuela:2016yny,Valenzuela:2017bvg,Blumenhagen:2017cxt,Montero:2017yja,Farakos:2017ocw,Lanza:2019xxg},
supersymmetric field
theories~\cite{Gates:1980az,Buchbinder:1988tj,Binetruy:1995hq,Binetruy:1995ta,Binetruy:1996xw,Ovrut:1997ur,Dvali:1999pk,Kuzenko:2005wh,Hartong:2009az,Seiberg:2010qd,Groh:2012tf,Farakos:2016hly,Becker:2016xgv,Aoki:2016rfz,Yokokura:2016xcf,Farakos:2017jme,Kuzenko:2017vil,Buchbinder:2017vnb,Kuzenko:2017oni,Bandos:2018gjp,Cribiori:2018jjh,Nitta:2018vyc,Nitta:2018yzb,Bandos:2019qok,Bandos:2019wgy}
(see Refs.~\cite{Gates:1983nr,Buchbinder:1998qv,Binetruy:2000zx} as a
review).  
We show that the topological charge of the domain walls is coupled with
the 3-form gauge field in a low-energy limit.
This dual 3-form gauge theory can be described by a topological
action of the 3-form gauge field and the original scalar field.
Then, we show that there are spontaneously broken ${\mathbb Z}_k$ 0-form
and ${\mathbb Z}_k$ 3-form symmetries, with nontrivial commutation
relation between them, thereby characterized by the type-B symmetry
breaking \cite{Watanabe:2011ec,Watanabe:2012hr,Hidaka:2012ym,
Watanabe:2014fva, Hayata:2014yga,Takahashi:2014vua,Watanabe:2019xul} as
a natural extension of s-wave superconductors.
We further discuss similar and different points between the phase
admitting the domain walls and topologically ordered phases.

This paper is organized as follows.
In section \ref{eff}, we give a low-energy effective theory of a system
of a periodic scalar field that admits domain walls.
In section \ref{dual}, we dualize the low-energy effective theory to a
3-form gauge theory with a topological coupling between the 3-form gauge
field and the original periodic scalar field.
We further show that the low-energy limit is described by the
topological field theory with the topological coupling.
By using the dual topological field theory, we show in section
\ref{higher} that there is an emergent $\bb{Z}_k$ 3-form global
symmetry whose charged object is a domain wall worldvolume, and the
3-form symmetry is spontaneously broken.
We further show that the symmetry breaking pattern is classified as a
type-B spontaneous symmetry breaking.
In section \ref{top}, we discuss similar and different points between
the phase with domain walls and topologically ordered phases.
In section \ref{phi4}, we comment on an emergent 3-form symmetry in the
$\Phi^4$ model.
Finally, we summarize this paper in section \ref{sum}.

\section{Effective theory of domain wall}
\label{eff}
Here, we show how the system with domain walls
can be described in low-energy effective theories.
First, we consider a theory that admits a domain wall solution.
We introduce a scalar field $\phi$ which obeys the $2\pi$ periodicity:
\begin{equation}
\phi({\cal P}) + 2\pi \sim \phi({\cal P}), 
\label{191201.2024}
\end{equation}
where ${\cal P}$ denotes a point in the spacetime.
Note that the mass dimension of the scalar field is normalized as 0.
Since adding $2\pi$ to $\phi$ is redundant, the 
physical observables made of $\phi$ are functions of $e^{iq\phi}$ 
with $q \in \bb{Z}$
in order to make them invariant under the redundant transformation in \er{191201.2024}.

Now, we give an action which admits a stable domain wall\footnote{
We use differential form notation: $d$ and $\star$ denote the exterior differential
and Hodge's star operator, respectively.
In particular, the density $4$-form $d^4 x$ can be given by $\star 1$.}:
\begin{equation}
 S = -\int (\tf{v^2}{2}|d\phi|^2 +  V(\phi) \star 1).  
\end{equation}
Here, $v$ is a mass-dimension 1 parameter.
The potential 
$V(\phi)$ is a periodic function $V(\phi+ {2\pi} / {k}) = V(\phi)$
with $k \in \bb{Z}$.
The minima of $V$ are given by $\phi = {2\pi n}/{k}$ with 
$V({2\pi n} / {k}) =0$, 
$V'({2\pi n} / {k}) =0$, 
and
$ V''({2\pi n} / {k})  >0$, 
where $n \in \bb{Z}$ mod $k$.
An example of $V(\phi)$ can be $V(\phi) = V_0(1- \cos k\phi)$,
which is known as a sine-Gordon model,
but we do not specify the detail of the potential.

The action has a $\bb{Z}_k$ global symmetry of the scalar field,
$e^{i\phi} \to \omega e^{i\phi} $,
where $\omega\in \bb{Z}_k$.
This discrete symmetry is spontaneously broken.
Domain walls are introduced as configurations,
which connect different minima.

In the presence of the domain walls, the topological charge of the domain walls
can be given by
\begin{equation}
\begin{split}
 Q({\cal P}_{\infty},{\cal P}'_{\infty}) 
= &\fr{k}{2\pi}(\phi({\cal P}_{\infty}) - \phi ({\cal P}_{\infty}'))
=
\fr{k}{2\pi}
\int^{{\cal P}_{\infty}}_{{\cal P}_{\infty}'} 
d \phi,
\end{split}
\label{191130.0308}
\end{equation}
where ${\cal P}_{\infty}$ and ${\cal P}'_{\infty}$ are points at infinity.
Because of the redundancy in \er{191201.2024}, the topological charge 
should be $ Q({\cal P}_{\infty},{\cal P}'_{\infty})  \in \bb{Z}$ mod $k$ 
in other words, 
\begin{equation}
 e^{\fr{2\pi i}{k} Q({\cal P}_{\infty},{\cal P}'_{\infty})}
 = \exp
\(i\int^{{\cal P}_{\infty}}_{{\cal P}_{\infty}'} 
d \phi
\) 
\label{191204.2347}
\end{equation} 
is physical.
Note that the integration of $d\phi$ around a closed loop ${\cal C}$,
$\int_{\cal C} d\phi$ 
is invariant under the identification in \er{191201.2024}.
The topological nature of $Q({\cal P}_{\infty},{\cal P}'_{\infty}) $
is that $Q({\cal P}_{\infty},{\cal P}'_{\infty}) $ 
is not changed under small deformations of ${\cal P}_{\infty}$ or 
${\cal P}_{\infty}'$ which do not pass through the domain walls.

Now, we consider the low-energy limit in order to see the phase structure of the system.
Generally, domain walls have a finite width.
However, the width can be neglected in the low-energy (long-range) limit.
To describe low-energy dynamics, we split $\phi$ into 
the domain wall part $\phi_{\rm W}$ and the fluctuation part $\phi_{\rm F}$:
$\phi = \phi_{\rm W} + \phi_{\rm F}$.
We assume the boundary condition of $\phi_{\rm W}$ and $\phi_{\rm F}$ are
\begin{equation}
 \phi_{\rm W}|_{\infty} = \phi|_{\infty},
\quad
 \phi_{\rm F}|_{\infty} =0,
\label{191205.1707}
\end{equation}
where `$|_{\infty}$' stands for the value at infinity.

In order to describe domain walls in the low-energy effective theory,
$\phi_{\rm W}$ is introduced as
\begin{equation}
 d\phi_{\rm W} = \fr{2\pi}{k} \sum_i \delta_1({\cal V}_i).
\end{equation}
Here, ${\cal V}_i$ ($i = 1,2,3,...$) denotes 
a 3-dimensional subspace of the 
core of the $i$-th domain wall.
The $\delta_1 ({\cal V}_i)$ is a delta function 1-form 
given by
\begin{equation}
\begin{split}
 \delta_{1}({\cal V}_i) 
= \fr{\epsilon_{mnpq}}{3!} 
dx^q
\int_{{\cal V}_i} 
dy^{m} \wed dy^n \wed dy^p
\delta^4 (x-y).
\end{split}
\end{equation}
Since the domain walls are described by the  
delta function, 
the topological charge in \er{191204.2347} can be 
described by two points in the spacetime ${\cal P}$ and ${\cal P}'$
except for the core of the domain walls: 
\begin{equation}
\begin{split}
&  e^{ \fr{2\pi i }{k} Q({\cal P},{\cal P}')}
 =  
e^{i \phi_{\rm W} ({\cal P}) - i\phi_{\rm W} ({\cal P'})}
\\
&
 =  
e^{i \int_{{\cal P'}}^{{\cal P}} d\phi_{\rm W}}
=
e^{\fr{2\pi i}{k}
\sum_i \link (({\cal P},{\cal P}'), {\cal V}_i)}
\end{split}
\end{equation}
Here, 
$\link (({\cal P},{\cal P}'), {\cal V}_i)$
denotes the linking number of 
the two points 
$({\cal P},{\cal P}')$
and ${\cal V}_i$.
The ``linking'' of 
$({\cal P},{\cal P}')$
and ${\cal V}_i$ means that 
we cannot move ${\cal P}$ continuously to ${\cal P}'$ 
without passing through ${\cal V}_i$.
The domain wall part
$\phi_{\rm W}$ consistently gives us the topological charge in \er{191130.0308}
because of the boundary conditions in \er{191205.1707}.

In the presence of the domain walls, the 
low-energy effective action can be given by the fluctuations around the 
configuration of the domain walls.
Up to the second order of fluctuation $\phi_{\rm F} = \phi - \phi_{\rm W}$,
the low-energy effective action is given by 
\begin{equation}
 S_{\rm eff} 
= -\int \(\fr{v^2}{2}|d\phi|^2 +\fr{\lambda^4}{2} (\phi -\phi_{\rm W})^2 \star 1\),
\label{191203.1637}
\end{equation}
where we have defined $\lambda^4 = V''({2\pi n } / {k})$.
\section{Dual 3-form formulation}
\label{dual}
In order to show the topological aspects,  
we dualize the action in \er{191203.1637} 
into a 3-form gauge theory.
We can rewrite $S_{\rm eff}$ by introducing a Lagrange's multiplier 
$c_3$ and an auxiliary scalar $f$ with $2\pi$ periodicity $f +2\pi \sim f$: 
\begin{equation}
\begin{split}
 S_{\rm eff,1st} 
=& - \int \(\fr{v^2}{2} |d\phi|^2 + \fr{\lambda^4}{2} (\phi - f)^2 \star 1\) 
\\
&
+ \fr{k}{2\pi} \int c_3 \wed d(f - \phi_{\rm W}), 
\end{split}
\label{191203.1651}
\end{equation}
with the boundary condition
$f|_{\infty} =  \phi_{\rm W}|_{\infty}$.
Note that the normalization of $c_3$ is determined so that 
the coupling of $c_3$ and the domain walls are normalized.
The action $ S_{\rm eff,1st} $ is invariant under the gauge transformation of
the 3-form: $c_3 \to c_3 + d\lambda_2$ 
with an arbitrary 2-form $\lambda_2$. 

The equation of motion for $c_3 $ with the boundary condition 
leads to $f= \phi_{\rm W}$, and we return to the original action $S_{\rm eff}$.
Instead, if we eliminate $f$ by substituting its equation of motion,
$f = \phi - {k} / ({2\pi \lambda^4})  \star dc_3$,
into $S_{\rm eff,1st}$, we reach the dual action 
with domain walls:
\begin{eqnarray}
  S_{\rm dual}
 = 
&-\ds{\int} \(\fr{v^2}{2}|d\phi|^2 + \fr{k^2}{8\pi^2 \lambda^2}|dc_3|^2
-
\fr{k}{2\pi} 
 c_3 \wed d\phi 
\)  
\nonumber
\\
&
 + \df{k^2}{4\pi^2 \lambda^2}\ds{\int} d(c_3 \wed  \star  dc_3)
+\sum_i \int_{{\cal V}_i}c_3.  
\label{191204.1710}
\end{eqnarray}
Here, the term $|dc_3|^2$ is the kinetic term of the 3-form gauge field,
and the term $d(c_3 \wed  \star dc)$ is 
the boundary term for the kinetic term.
Note that the boundary term is generally necessary in order to 
have an energy-momentum tensor consistent 
with 
the equation of motion for 
$c_3$~\cite{Brown:1987dd,Brown:1988kg,Duff:1989ah,Duncan:1989ug,Duncan:1990fr,Nitta:2018yzb,Nitta:2018vyc}.
 
The dual transformation naturally gives us the electric coupling of 
the domain walls to the 3-form gauge field.
We can identify the domain wall worldvolume operator by the dual action as
$ \exp(i\int_{\cal V} c_3)$.

In the low-energy limit,
we can neglect the kinetic terms of the scalar field and 
the 3-form gauge field as well as the boundary term for the kinetic term
since the system is gapped.
Therefore, the system can be described by 
the topological action
\begin{equation}
 S_{\rm top.}
= \fr{k}{2\pi} \int c_3 \wed d\phi, 
\end{equation}
together with the domain wall worldvolume operators.
We can anticipate that the symmetry breaking pattern is 
so-called type-B spontaneous symmetry braking,
since the topological action is given by a first order temporal derivative term,
$c_3 \wed d\phi \sim c_{123} \der_0 \phi $.
The topological action is quite similar to 
the $BF$ action~\cite{Horowitz:1989ng,Blau:1989bq}.
Thus, we expect that the system described by $S_{\rm top.}$ has 
topologically non-trivial properties,
since the $BF$ action 
describes a topologically ordered phase, e.g., superconductors~\cite{Hansson:2004wca}.

\section{Higher-form symmetry breaking}
\label{higher}

To show the symmetries of the system and their breaking, 
we evaluate a correlation function 
of the topological action.
In the quantum theory, the observables are one-point operators
and the domain wall worldvolume operators
\begin{equation}
 I_0 (q;{\cal P}) =  e^{i q \phi ({\cal P})},
\quad
 D(p;{\cal V}) = e^{ip \int_{\cal V} c_3},
\end{equation}
respectively.
Here, ${\cal P}$ denotes a point in the spacetime.
Both of the charges $p$ and $q$ are integers,
since 
the charge $p$ of the domain wall worldvolume operator is 
quantized by the topological charge,
and the charge $q$ of the one-point operator is quantized due
to the $2\pi $ redundancy in \er{191201.2024}.
In order to capture the topological charge of the domain walls, 
 we introduce the following two-point operator
\begin{equation}
 I(q;{\cal P}, {\cal P}') = e^{iq\phi ({\cal P}) -i q\phi ({\cal P}')}
\end{equation}
as in \er{191130.0308} rather than $I_0(q;{\cal P})$.
In general, it is difficult to say whether there is a domain wall or not 
by using only one point operator $I_0(q; {\cal P})$, since 
domain walls are generally non-compact, and do not surround the one point operator.

The correlation function can be evaluated by using $S_{\rm top.}$ as
\begin{equation}
 \vevs{I(q;{\cal P,P'}) D(p;{\cal V})} 
= e^{-\fr{2\pi i qp}{k}\link (({\cal P, P'}),{\cal V} )}.
\label{191130.0318}
\end{equation}
One can also show the VEVs of the 
domain wall worldvolume and two-point operators are non-zero:
\begin{equation}
 \vev{D(p;{\cal V})} = \vev{I(q;{\cal P,P'})} =1.
\label{191130.0320}
\end{equation}
The relation in \er{191130.0318} shows that 
there is a fractional linking phase 
between the domain wall worldvolume and two-point operators.
Furthermore, \er{191130.0320} means that the VEVs of the non-local objects are 
non-zero.

The fractional linking phase as well as non-zero VEVs of the 
non-local operators in the long range (low energy) limit can be seen as 
spontaneously broken $\bb{Z}_k$ 0- and 3-form symmetries.
For the 0-form symmetry, the charged object is $I(q;{\cal P,P'})$ 
and the symmetry generator is $D(p;{\cal V})$.
The $\bb{Z}_k$ transformation is generated by 
\begin{equation}
  \vevs{ D(p;{\cal V})I(q;{\cal P,P'})} = 
e^{-\fr{2\pi i qp}{k}\link (({\cal P, P'}),{\cal V} )} \vevs{I(q;{\cal P,P'})}.
\end{equation}
For the 3-form symmetry, the charged object is $D(p;{\cal V})$ 
and the symmetry generator is $I(q;{\cal P,P'})$.
The $\bb{Z}_k$ transformation is generated by 
\begin{equation}
  \vevs{ I(q;{\cal P,P'}) D(p;{\cal V})} = 
e^{-\fr{2\pi i qp}{k}\link (({\cal P, P'}),{\cal V} )} \vevs{D(p;{\cal V})}.
\end{equation}
This 3-form symmetry is an emergent symmetry in the low-energy 
effective theory.
Both of $D(p;{\cal V})$ and $I(q;{\cal P,P'})$ are topological
because of the relation in \er{191130.0320}.
Therefore,
both of 0- and 3-form symmetries are broken spontaneously, 
since the charged objects develop non-zero VEVs as in \er{191130.0320}.

This symmetry breaking can also be seen in the type-B spontaneous symmetry
breaking viewpoint.
First, the breaking of the 0- and 3-form symmetries can be classified into
the type-B spontaneous symmetry breaking, since the order parameters are symmetry generators.
Second, we can choose the ground state as an eigenstate of the 
symmetry generators, that is, the order parameters.
Since the symmetry generators are unitary operators, the 
eigenvalues of the charged objects are non-zero.
Therefore, the VEVs of the order parameters are finite.

\section{Comparison to topological ordered phases}
\label{top}

We have seen that 
there exist spontaneously broken discrete 0- and 3-form symmetries,
and the charged objects are the symmetry generators.
The symmetry breaking pattern is quite similar to the ones in
topologically ordered phases in fractional quantum Hall effects (FQHE)
in $(2+1)$ dimensions and s-wave superconductors in $(3+1)$ dimensions.
In those systems, the charged objects are also symmetry generators.
Therefore, we conclude that the phase admitting the domain walls is a
generalization of the topologically ordered phase.

However, there are different points between our system and the 
previously known topologically ordered phases due to the dimensions of the 
charged objects.
One is the ground state degeneracy on a compact spatial manifold.
On one hand, the FQHE and s-wave superconductors lead to ground state
degeneracy depending on the topology of the compact manifold, since the
Wilson loop and vortex surface operators can capture the non-trivial
cohomology classes of the compact manifolds.
On the other hand, the domain wall system always leads to ground state
degeneracy because the domain wall worldvolume can always wrap the
three-dimensional compact orientable manifold.

Another different point is whether the braiding phase exists or not.
In the FQHE and s-wave superconductors, there can be braids of the
charged objects that represent trajectories of anionic excitations,
since the charged objects are extended objects.
However, in the domain wall case, there cannot be braids of the charged
objects, because the two-point operators are not extended objects.

\section{Other domain walls}
\label{phi4}
In the above discussions, we have considered the domain walls made of 
the periodic scalar field.
However, domain walls can arise without periodicity, 
e.g., $\Phi^4$ model.
We will show that there is also an emergent 3-form discrete symmetry
spontaneously broken in the gapped phase.

We consider 
a Lagrangian of a real scalar field $\Phi$ 
\begin{equation}
 S_\Phi = -\int\(\fr{1}{2} |d\Phi|^2 + \fr{\alpha}{4}(\Phi^2- v^2)^2\) 
\star 1
\end{equation}
with a positive parameter $\alpha$.
The action has a $\bb{Z}_2$ 0-form symmetry under the 
transformation $\Phi \to -\Phi$.
In the spirit of the generalized global symmetry, the existence of the 
$\bb{Z}_2$ 0-form symmetry implies the existence of 
a topological operator $U(\sigma, {\cal V})$ parameterized by 
$\sigma \in \bb{Z}_2 =\{1,-1\}$ and a 3-dimensional closed surface ${\cal V}$.
The transformation law is given by 
\begin{equation}
 \vevs{U(\sigma, {\cal V}) \Phi({\cal P})\Phi({\cal P'})} 
= \sigma^{\link (({\cal P,P'}),{\cal V})} \vevs{\Phi({\cal P})\Phi({\cal P'})}.
\end{equation}

In the $\Phi^4$ model, there are two vacua characterized by
 $\vev{\Phi} = \pm v $ and $\vev{d \Phi} =0$.
Since the vacua are gapped, the correlation function of $\Phi$,
$\vevs{\Phi({\cal P})\Phi({\cal P'})} = v^2$, 
and $\vevs{(d\Phi({\cal P})) \Phi({\cal P'})} =0$ in the long range limit.
Therefore, the two-point operator $\Phi({\cal P})\Phi({\cal P'})$ is a topological 
operator.
This implies that the two-point operator $\fr{1}{v^2}\Phi({\cal P})\Phi({\cal P'})$
becomes a symmetry generator of a $\bb{Z}_2$ 3-form global symmetry:
the charged object is $U(\sigma, {\cal V})$, and the 
transformation law is a $\bb{Z}_2$ transformation given by
\begin{equation}
 \fr{1}{v^2}\vevs{U(\sigma, {\cal V}) \Phi({\cal P})\Phi({\cal P'})} 
= \sigma^{\link (({\cal P,P'}),{\cal V})} \vevs{U(\sigma, {\cal V})},
\end{equation}
where ${\cal P,P'}$ and ${\cal V}$ are sufficiently separated.
Since $\vevs{U(\sigma, {\cal V})} = \fr{1}{v^2} \vevs{\Phi({\cal
P})\Phi({\cal P'})} =1$, the 3-form global symmetry is spontaneously
broken.

Ferromagnets with easy axis potential allow
magnetic domain walls, which can be described in the massive ${\mathbb C}P^1$ model with ground states at north and south poles (see, e.g., \cite{Nitta:2012xq}).  
If we restrict the trajectory to a great circle, the model is reduced to
the sine-Gordon model with $k=2$.  
It is open question whether the massive ${\mathbb C}P^1$ model, 
${\mathbb C}P^{N-1}$ or Grassmannian generalization 
\cite{Eto:2006pg} exhibits 
topological orders.

\section{Summary}
\label{sum}

We have shown that there is an emergent $\bb{Z}_k$ 3-form global
 symmetry spontaneously broken in a gapped phase admitting domain walls.
The symmetry generator and the charged object for the 3-form symmetry
are a two-point operator and a domain wall worldvolume operator,
respectively.

In order to show the existence of the 3-form symmetry and its breaking,
we have established the dual 3-form formulation of the massive periodic
scalar field.
In the dual formulation, we have evaluated the correlation function of
the two-point and domain wall worldvolume operators.
The correlation functions can be characterized by a fractional linking
phase, and both the domain wall worldvolume and two-point operators
develop finite VEVs.

The phase can be characterized by a spontaneous breaking of the
$\bb{Z}_k$ 0-form and 3-form global symmetries with nontrivial
commutation relation between them, thereby characterized by the
so-called type-B symmetry breaking.
This phase is similar to the conventional topologically ordered phase in
the sense of the symmetry breaking pattern as well as the fractional
linking phase, while the different points are quasi-particle excitations
and ground state degeneracy on a compact spatial manifold, due to the
spatial dimensions of symmetry generators.
Obviously, our analysis can be generalized to $(d+1)$ dimensions, by
replacing the worldvolume operator $c_{3}$ by $c_{d}$, although we
focused on the axion models in $(3+1)$ dimensions.

There are several avenues for future work.
In particular, our result has an implication to a domain wall problem in
axion cosmology, since topologically non-trivial case, $k >1$,
corresponds to the case admitting stable axion domain walls causing the
domain wall problem.
For that purpose, a dual transformation including axion strings as well
as axion domain walls is needed, which we address elsewhere.

\paragraph*{Note added.} 

While this work was being completed, we received
Ref.~\cite{Tanizaki:2019rbk} where spontaneous breaking of a discrete
3-form global symmetry was considered in Yang--Mills theories.

\subsection*{Acknowledgements}
R.~Y.~thanks Toshiaki Fujimori, Tatsuhiro Misumi, 
and Kazuya Yonekura
for helpful discussions.
The discussions during the KEK Theory Workshop 2019 
were useful in completing this work.
The work of Y.~H.~is supported in part by 
Japan Society of Promotion of Science (JSPS) 
Grant-in-Aid for Scientific Research 
(KAKENHI Grants No.~16K17716, 17H06462, and 18H01211), by RIKEN iTHEMS.
This work of M.~N.~was supported by the 
Ministry of Education, Culture, 
Sports, Science (MEXT)-Supported Program for the Strategic Research 
Foundation at Private Universities ``Topological  Science'' 
(Grant  No.~S1511006).
This work of M.~N.~is also supported 
in part by JSPS KAKENHI Grant Numbers 16H03984, and 18H01217.
The work of M.~N.~is also supported in part by 
a Grant-in-Aid for Scientific Research on Innovative Areas 
``Topological Materials Science'' 
(KAKENHI Grant No.~15H05855) from MEXT of Japan.

\end{document}